\documentclass[sigconf]{acmart}
\usepackage{algpseudocode}
\usepackage{url}
\usepackage{balance}
\usepackage{csvsimple}
\usepackage{etoolbox}
\usepackage{tabularx}
\usepackage{wasysym}
\usepackage{algorithm}
\usepackage{booktabs}
\usepackage{enumerate}
\usepackage{multicol}
\usepackage{mdframed}
\usepackage{multirow}
\usepackage{url}
\usepackage{comment}
\usepackage{xcolor,colortbl}
\usepackage{color,soul}
\usepackage{pifont}
\usepackage{multirow}
\usepackage{caption}
\usepackage{makecell}
\usepackage{subfig}
\usepackage{graphicx}
\newcolumntype{?}{!{\vrule width 1pt}}
\usepackage{listings}
\usepackage[title]{appendix}
\usepackage{appendix}
\usepackage{float}

\definecolor{mGreen}{rgb}{0,0.6,0}
\definecolor{mGray}{rgb}{0.5,0.5,0.5}
\definecolor{mPurple}{rgb}{0.58,0,0.82}
\definecolor{backgroundColour}{rgb}{0.95,0.95,0.92}
\definecolor{notecolor}{rgb}{0,0.0,0} 

\lstdefinestyle{CStyleListing}{
	commentstyle=\color{mGreen},
	keywordstyle=\color{magenta},
	numberstyle=\tiny\color{mGray},
	stringstyle=\color{mPurple},
	basicstyle=\footnotesize\ttfamily,
	breakatwhitespace=false, 
	breaklines=true,                 
	captionpos=b,                    
	keepspaces=true,                 
	numbers=left,                    
	numbersep=3pt,                  
	showspaces=false,                
	showstringspaces=false,
	showtabs=false,                  
	tabsize=4,
	aboveskip=2mm,
	belowskip=2mm,
	language=c,
	xleftmargin=0.1in,
	xrightmargin=0.2in
}

\definecolor{codegreen}{rgb}{0,0.6,0}
\definecolor{codegray}{rgb}{0.5,0.5,0.5}
\definecolor{codepurple}{rgb}{0.58,0,0.82}
\definecolor{backcolour}{rgb}{0.95,0.95,0.92}

\lstdefinestyle{mystyle}{
    backgroundcolor=\color{backcolour},   
    commentstyle=\color{mGreen},
    keywordstyle=\color{magenta},
    numberstyle=\tiny\color{codegray},
    stringstyle=\color{codepurple},
    basicstyle=\fontsize{6.5}{7.5}\ttfamily,
    breakatwhitespace=false,         
    breaklines=true,                 
    captionpos=b,                    
    keepspaces=true,                 
    numbers=left,                    
    numbersep=5pt,                  
    showspaces=false,                
    showstringspaces=false,
    showtabs=false,                  
    tabsize=2
}

\newcommand{\etal}{\emph{et al.}}
\newcommand{\eg}{\emph{e.g.}}
\newcommand{\ie}{\emph{i.e.}}
\definecolor{aogreen}{rgb}{0.0, 0.5, 0.0}
\newcommand{\song}[1]{\textcolor{black}{#1}}
\newcommand{\liao}[1]{\textcolor{black}{#1}}
\newcommand{\ls}[1]{\textcolor{black}{#1}}
\newcommand{\songl}[1]{\textcolor{black}{#1}}
\newcommand{\cmark}{\ding{51}}%
\newcommand{\xmark}{\ding{55}}%

\newboolean{showcomments}
\setboolean{showcomments}{false}
\ifthenelse{\boolean{showcomments}}
{\newcommand{\hc}[1]{{\color{magenta} Haipeng: #1}}}{\newcommand{\hc}[1]{}}

\newcommand{\ProjectName}[1]{{\small\textsc{SkillScanner}}}
\newcommand{\clnote}[1]{\textcolor{blue}{#1}}

\copyrightyear{2023}
\acmYear{2023}
\setcopyright{rightsretained}
\acmConference[CCS '23]{Proceedings of the 2023 ACM SIGSAC Conference on Computer and Communications Security}{November 26--30, 2023}{Copenhagen, Denmark}
\acmBooktitle{Proceedings of the 2023 ACM SIGSAC Conference on Computer and Communications Security (CCS '23), November 26--30, 2023, Copenhagen, Denmark}\acmDOI{10.1145/3576915.3616650}
\acmISBN{979-8-4007-0050-7/23/11}

\begin{document}

%
\title{SkillScanner: Detecting Policy-Violating Voice Applications Through Static Analysis at the Development Phase}


\author{Song Liao}
\affiliation{%
 \institution{Clemson University}
 \country{United States}
}
\email{liao5@g.clemson.edu}

\author{Long Cheng}
\affiliation{%
 \institution{Clemson University}
 \country{United States}
}
\email{lcheng2@clemson.edu}

\author{Haipeng Cai}
\affiliation{%
 \institution{Washington State University}
 \country{United States}
}
\email{haipeng.cai@wsu.edu}

\author{Linke Guo}
\affiliation{%
 \institution{Clemson University}
 \country{United States}
}
\email{linkeg@clemson.edu}

\author{Hongxin Hu}
\affiliation{%
 \institution{University at Buffalo}
 \country{United States}
}
\email{hongxinh@buffalo.edu}



\begin{abstract}

The Amazon Alexa marketplace is the largest Voice Personal Assistant (VPA) platform with over 100,000 voice applications (\ie, skills) published to the skills store. In an effort to maintain the quality and trustworthiness of voice-apps, Amazon Alexa has implemented a set of policy requirements to be adhered to by third-party skill developers. However, recent works reveal the prevalence of policy-violating skills in the current skills store.
To understand the causes of policy violations in skills, we first conduct a user study with 34 third-party skill developers focusing on whether they are aware of the various policy requirements defined by the Amazon Alexa platform. Our user study results show that there is a notable gap between VPA's policy requirements and skill developers' practices. As a result, it is inevitable that policy-violating skills will be published.

To prevent the inflow of new policy-breaking skills to the skills store from the source, it is critical to identify potential policy violations at the development phase.
In this work, we design and develop \ProjectName{}, an efficient static code analysis tool to facilitate third-party developers to detect policy violations early in the skill development lifecycle.
To evaluate the performance of \ProjectName{}, we conducted an empirical study on \song{2,451} open source skills collected from GitHub. 
\ProjectName{} effectively identified \song{1,328} different policy violations from \song{786} skills. Our results suggest that 32\% of these policy violations are introduced through code duplication (\ie, code copy and paste). In particular, we found that 42 skill code examples from potential Alexa's official accounts \liao{(\eg, ``alexa'' and ``alexa-samples'' on GitHub)} contain policy violations, which lead to 81 policy violations in other skills due to the copy-pasted code snippets from these Alexa's code examples.



\end{abstract}


\maketitle

%

\section{Introduction}
Amazon Alexa is one of the leading Voice Personal Assistant (VPA) platforms that allow third-party developers to build new voice applications (\ie, called skills) and publish them to the skills store~\cite{FirstAlexaSkill}. 
The openness of the Amazon Alexa platform has greatly attracted skill developers and inflated VPA's capabilities.
Currently, more than 100,000 Alexa skills are available~\cite{SkillsCount2022}, with functions such as ordering pizza, listening to the news and weather, locking doors, or checking credit card balance. 
However, such an open VPA ecosystem inevitably provides unscrupulous or inexperienced developers an opportunity to publish buggy or dangerous skills in the store~\cite{DangerousSkills:2019, Long:2020:CCS}. Consequently, these poor-quality and problematic skills could cause user frustration, stir up negative effects on engagement, and even place end users in a vulnerable position.


To ensure the content safety and privacy of skills in
Alexa skills store, Amazon has defined various policy requirements, including 7 privacy requirements~\cite{PrivacyRequirements}, 14 main sections of content guidelines~\cite{PolicyTesting}, \liao{and code inconsistency~\cite{AlexaTesting}}
to be adhered to by third-party skill developers.
For example, skills should not have advertisements or promote alcohol.
Although these policies are checked during the skill certification process (which rejects a skill if it violates any of the pre-defined policies), prior work demonstrated the ease of policy-violating skills being certified~\cite{Long:2020:CCS, Wang:Ubicom:2021}. Several recent works~\cite{SkillExplorer:2020,Shezan:VerHealth:2020, Lentzsch:NDSS:2021,li2022vitas, SkillDetective:2022} developed tools to measure the policy compliance of skills on the Amazon Alexa platform through a dynamic analysis approach (\ie, by exploring the outcomes of skills). 
For example, SkillExplorer~\cite{SkillExplorer:2020} found 1,141 skills that collect different types of private information but without disclosing the data collection in their privacy policies. 

In particular, researchers in SkillDetective~\cite{SkillDetective:2022} found 3,473 Alexa skills violating the same policy whereby skills are forbidden to ``explicitly request that users leave a positive rating of the skill''. 
It is likely that many third-party developers were not completely aware of such policy requirements, and inadvertently violated them in their skills.  
Past research has shown that software developers find it difficult to understand various policies and requirements when they develop software applications~\cite{Swapneel:2014:ICSE, Awanthika:2018:EASE}, and are often unaware of all the related policies~\cite{Lisa:2017:BMC}.
In this work, we are curious about whether developers are aware of the various policy requirements defined by the VPA platform.
To this end, we first conduct an in-depth user study to understand third-party skill developers' perceptions and interpretations regarding VPA's policy requirements.
Our user study results show that there exists a notable gap between VPA's policy requirements and skill developers' practices.
To prevent the inflow of new policy-violating skills to the public, it is critical to identify potential policy violations in skills early in the skill development lifecycle.





\begin{figure*}[!h]
	\begin{center} 
		\includegraphics[width=0.85\textwidth]{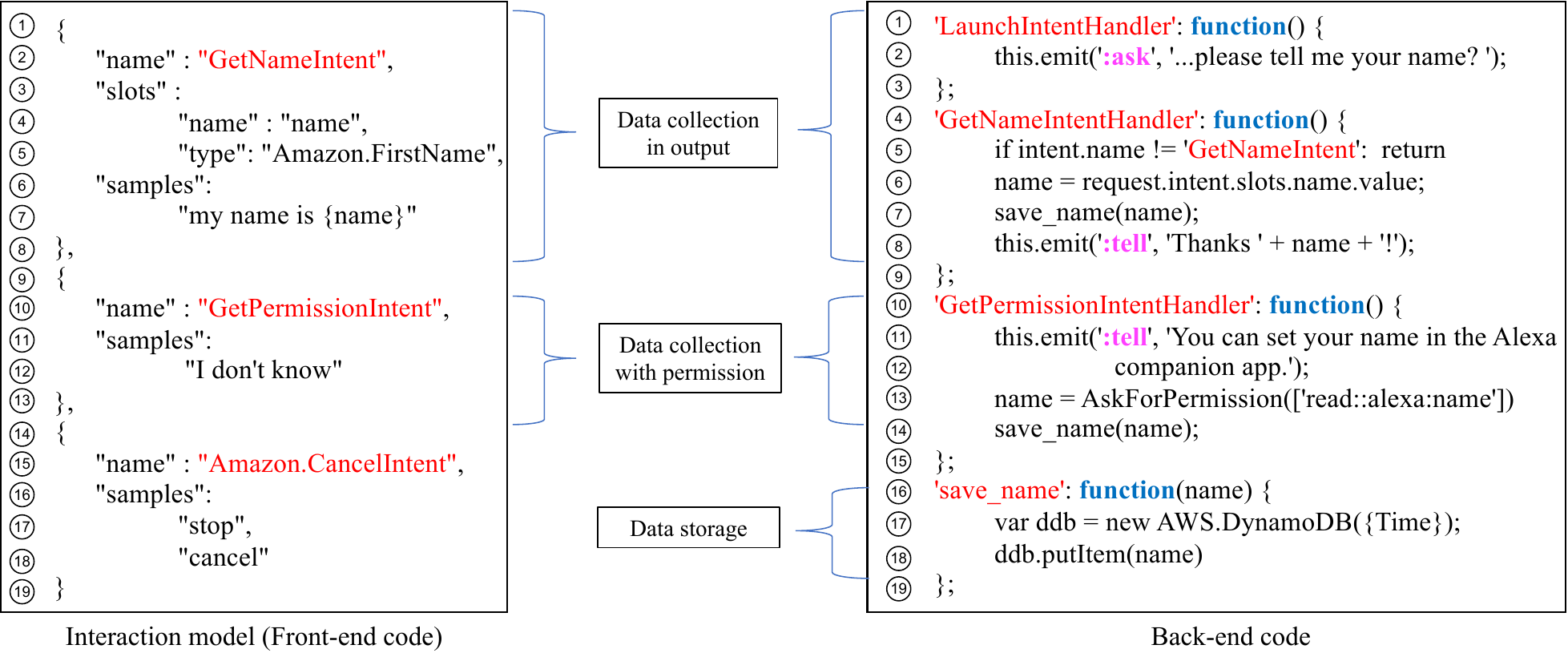}
	\end{center} 
	\vspace{-8pt}
	\caption{Example of the front-end and back-end code of an Alexa skill.}\label{code1}
	\vspace{-10pt}
\end{figure*}



In this work, we seek to develop static analysis techniques to facilitate the development of policy-compliant skills by third-party developers. 
However, the unique code structure of skills poses challenges for performing a static analysis of the skill code since we need to consider the interaction between the front-end code and back-end code during our analysis.
The diverse nature of policy requirements defined by Amazon Alexa is another challenge for detecting policy violations in skill code. 
To address these challenges, we propose \ProjectName{} to automatically evaluate skills' conformity to various policy requirements before their deployment. 
Compared with existing dynamic analysis works~\cite{SkillExplorer:2020, Shezan:VerHealth:2020, SkillDetective:2022},
\ProjectName{} takes a static analysis approach and  
is expected to identify more code-specific violations that the dynamic analysis approach cannot find.
In summary, we make the following contributions:
\begin{itemize} 
	\vspace{-2pt}
	\item To our best knowledge, \ProjectName{} is the first static analysis tool to facilitate the development of policy-compliant skills by third-party developers. 
	\ProjectName{} is able to detect various policy violations and code defects, which can effectively help developers improve the quality of their code, for a sustainable VPA ecosystem. We shared the \ProjectName{} tool, related datasets and results with the community to facilitate future research\footnote{The details of our tool implementation, evaluation results and representative skills are available at \url{https://github.com/CUSecLab/SkillScanner}.}.
	
	
	\item To understand the root causes of the prevalence of policy-violating skills in the skills store, we conducted a user study with 34 third-party skill developers. The results suggest that most skill developers in our study were not completely aware of the policy requirements defined by Amazon Alexa.  
	
	
	\item We collected a benchmark dataset for Alexa skill code with 2,451 open-source skills from the GitHub. We conducted a comprehensive analysis of these skills using \ProjectName{} and evaluated the performance of \ProjectName{}. We identified 1,328 different types of policy violations among 786 skills. 694 of them are about privacy issues and 298 skills violate the content guidelines defined by Amazon Alexa.
	32\% of these policy violations are because of code duplication (\ie, code copy\&paste). Table~\ref{table:results} summarizes our detection results. 
	Surprisingly, we found that 42 skill code examples from Alexa's official accounts contain policy violations, which led to 81 policy violations in other skills due to the code duplication. \liao{We discuss the responsible disclosure in Section~\ref{responsible}.}

\end{itemize}

\section{Background and Challenges}
\label{background}

\subsection{Skill Code Structure}
\label{CodeStructure}

A skill has a front-end interaction model (\ie, front-end code\footnote{According to Alexa Developer Documentation~\cite{SkillPackageFormat}, a skill package includes the skill's front-end interaction model file, back-end source code files, and skill manifest file. We \song{refer to} the front-end interaction model as front-end code in our analysis.
}) and back-end code that processes requests and tells a VPA device what to respond. Amazon Alexa Cloud provides hosting for the front-end interface of a skill~\cite{SkillInteractionModel}, but its back-end code can be hosted on the developer's server 
\song{(\eg, either hosted by AWS Lambda under the developer's account or other third-party servers).}
In addition, the skill code also contains a skill manifest file (\ie, named ``skill.json'')~\cite{SkillManifestSchema}, which stores the skill name, category, description, privacy policy, as well as permission information. Once a skill is certified and published in the Alexa's skills store, such information will be shown on the skill's web page.



{\bf Front-end code}. There are mainly three types of data defined in the front-end code: \textit{intent}, \textit{slot} and \textit{sample utterances}. 
An intent represents an action that fulfills a user's spoken request. \song{Intents normally have two values: the intent name and sample utterances.}
Sample utterances are a set of likely spoken phrases mapped to the intents and developers should include representative phrases so that the interaction model can better learn the sentence pattern.
In addition, intents can optionally have arguments called slots. Slot is the variable that can capture a specific type of user's verbal reply, such as username \song{or user address}.

When a skill is created, several default Amazon Alexa built-in intents are created automatically for basic functionality, \eg, \texttt{Amazon.StopIntent} for stopping a skill, and 
\song{\texttt{AMAZON.CancelIntent} for canceling an instruction}~\cite{AlexaBuiltinIntent}.
Developers can also create new intents by using an Alexa's pre-defined intent
or creating a custom intent. In the latter case, 
developers define the intent name and then provide some sample utterances so that when users give a reply semantically close to these utterances, the intent will be matched and invoked to process the user's request. 
\song{An intent can either be used for providing services (\eg, telling a story) or collecting data from users (\eg, \song{asking for names from users}).}
If developers want to use the intent to collect data from users, they need to define \song{slots} and specify the slot type. The slot type defines what data the slot collects, and developers can 
use Alexa's built-in slot types,
such as ``Amazon.FirstName'' trained with thousands of popular first names and ``Amazon.City'' for local and world cities.
For example, in Figure~\ref{code1}, \texttt{GetNameIntent} is a custom intent. It contains a slot ``\texttt{name}'' with the type ``Amazon.FirstName'', and the sample utterance is ``My name is \{name\}''. Then, when a user replies ``My name is Jack'', ``Jack'' will be extracted as a slot value and then passed to the back-end code.


{\bf Back-end code}. A skill's back-end code includes a list of intent handler functions for processing different intents defined in the front-end code. For example, \texttt{GetNameIntentHandler} in Figure~\ref{code1} is an intent handler corresponding to the intent \texttt{GetNameIntent} in the front-end code.
Several built-in intent handlers are provided by Amazon Alexa corresponding to the built-in intents, such as \texttt{HelpIntentHandler} for providing helpful information. 
\song{If no intent is triggered, the \texttt{FallbackIntentHandler} will be invoked and provide an output ``Sorry, I don't know about that. Please try again'' by default.}
Typically, an intent handler processes the received data,
performs skill functions and provides a reply to users.
In Figure~\ref{code1}, the \texttt{GetNameIntentHandler} retrieves the received name value from the user's verbal input, stores it in a local database, and generates an output to thank the user for providing the name.

\subsection{Challenges}
\indent{\bf The unique code structure of skills poses a challenge for data flow analysis. }
The back-end code of skill 
is tightly coupled with the front-end code. User input is first sent to the front-end interaction model, matched with intent, and then processed by the back-end code, as shown in Figure~\ref{code1}. 
In this example, there is a data collection request in line 2 of the back-end code.
From the front-end code, we know that
if users provide a name, it will be matched with the intent \texttt{GetNameIntent} (lines 2-8 in the front-end code) and then transferred to the function \texttt{GetNameIntentHandler} in the back-end code. Then, inside the function, 
the name data is stored in a local database and used for output (lines 16-18 and line 8 in the back-end code). The Amazon Alexa platform also provides APIs for skills to collect the information of device address, customer name, email address, phone number and location. In Figure~\ref{code1}, if the user doesn't provide a name (\eg, saying ``I don't know''), the \texttt{GetNamePermissionIntent} will be invoked and the skill will collect the user's name information using Alexa's built-in permission API \texttt{AskForPermission} in line 13 of the back-end code. \song{Due to the unique code structure of skills, existing static analysis tools cannot be directly applied to analyze the skill code since they are unable to adequately model the interaction between the front-end and back-end code. In \ProjectName{}, we develop static analysis techniques to extract connections between the front-end code and back-end code.}

\song{{\bf No existing benchmarks of skill code for static analysis research.} As discussed in Section~\ref{CodeStructure}, the back-end code of skills is hosted on the developer's server and developers don't need to submit the code for skill certification. 
For this reason, existing works only focused on the dynamic testing~\cite{SkillExplorer:2020, SkillDetective:2022, Shezan:VerHealth:2020} and analyzing the metadata of skills, such as permission information~\cite{edu2021skillvet, Jide:2022:WWW}, to identify potential issues in skills. 
We address this challenge by collecting a skill code dataset from the GitHub.
We have made our benchmark skill code dataset available to the community to facilitate future research.
}

\section{Understand Developers' Perception and Practice Regarding Policy Compliance}
\label{userstudy}

\song{We conduct a user study to understand the gap between VPA's policy requirements and skill developers' practices.} 
The user study has been approved by our university's IRB office.
We briefly describe the recruitment, survey questions, and results of our user study.

\subsection{Recruitment and Survey Questions}


There are skill developers leaving their emails in the skill descriptions for user feedback. We built a crawler to collect skill developers' contact emails from the US skills store and obtained 1,568 developer emails.
We used the Qualtrics platform~\cite{Qualitrics} to build survey questions and reach out to these developers to invite them to participate in our user study. 
{We included a pre-screening phase to ensure that the participants do have skill development experience.}
Finally, we received 44 responses but 10 participants' responses had to be removed due to the meaningless answers.
Our survey was conducted in August 2022, and we provided a \$10 Amazon gift card to each valid participant. The average time for completing the survey was 16 minutes. 
Although most of the participants are from the United States, there were \liao{a} few developers from other countries such as India and Italy. 10 participants developed more than 10 skills and the average number of skills developed by these participants is over 5, which indicates that they have enough coding experience in skill development. 8 participants also have experience of developing Google actions. 


Table~\ref{userstudy:general} lists several selected user study questions due to space. The complete survey questions and responses are available in our GitHub repository (\url{https://github.com/CUSecLab/SkillScanner}), which is composed of three sections. First, we asked developers whether they are aware of Amazon's policy requirements. 
Next, we asked developers whether they could identify any policy violation in skill outputs from 8 example skills (in which there are 5 policy violations in these skills).
Finally, we asked developers \song{whether they routinely perform} consistency checking during their skill development, such as code, content, privacy policy and whether they think that a static analysis tool can be helpful for them in developing policy-compliant skills.

\begin{table}[h]
	\centering
	\resizebox{8.5cm}{!}{
		{
			\begin{tabular}{  
			>{\arraybackslash}m{4cm} >{\arraybackslash}m{3.5cm} >{\raggedleft\arraybackslash}m{1.5cm}}
				\Xhline{5\arrayrulewidth}
				\rowcolor[gray]{0.9} \bf  Question & \bf Response & \bf Developers\\
				\hline
	
\multirow{5}{*}{\parbox{4cm} { Q1: Are you aware that there are some platform required policies?}}
    & Completely aware              &   65\% \\
	& Somewhat aware                & 	32\% \\
	& Neither aware nor unaware     & 	0\% \\
	& Somewhat unaware	            &   3\% \\	
	& Completely unaware	            &   0\% \\	
\hline			
\multirow{5}{*}{\parbox{4cm} {Q2: Do you read the \song{Alexa's policy requirements} before developing skills?}} 
    & Always                    &   24\% \\
	& Most of the time                 & 	26\% \\
	& About half the time       & 	32\% \\
	& Sometimes	                &   9\% \\	
	& Never	                    &   9\% \\	
\hline	



\multirow{5}{*}{\parbox{4cm} {\song{Q3: Do you think the Amazon Alexa platform should provide a policy compliance training for developers before they develop skills?}}} 
    & Definitely Yes            & 	29\% \\
	& Probably Yes              & 	47\% \\
	& It doesn't matter to me   &   6\% \\
	& Probably not	            &   18\% \\
	& Definitely not	        &   0\% \\
\hline			
		
\multirow{5}{*}{\parbox{4cm} {\song{Q4: Do you think the Amazon Alexa platform should check the skill code to determine whether it violates any policies?}}} 
    & Definitely Yes            &	24\% \\
	& Probably Yes              & 	47\% \\
	& It doesn't matter to me   & 	18\% \\
	& Probably not              & 	6\% \\
	& Definitely not            & 	6\% \\
\hline			
	
\multirow{5}{*}{\parbox{4cm} {Q5: Do you check whether your privacy policy and code are consistent?}} 
    & Always                    &   26\% \\
	& Most of the time                 & 	24\% \\
	& About half the time       & 	35\% \\
	& Sometimes	                &   3\% \\	
	& Never	                    &   12\% \\	
\hline	


\multirow{5}{*}{\parbox{4cm} {Q6: Do you fully inspect your code and remove any inconsistent code before submission?}} 
    & Always                    &   35\% \\
	& Most of the time                 & 	32\% \\
	& About half the time       & 	9\% \\
	& Sometimes	                &   18\% \\	
	& Never	                    &   6\% \\	
	
		\Xhline{5\arrayrulewidth}
			\end{tabular}
		}
	}
	\caption{Selected user study questions and responses.}
	\label{userstudy:general}
	\vspace{-25pt}
\end{table}

\subsection{Survey Results}

Although 33 skill developers (out of 34 participants) claimed that they are “completely aware” or “somewhat aware” that there are some platform required policies (Q1 in Table~\ref{userstudy:general}), most participants couldn't correctly recognize the 5 policy violations in our example skills (results are shown in Table~\ref{userstudy:skills}).
\song{When asked “Do you read Alexa's policy requirements before developing skills?” (Q2), only 8 developers selected ``always read'' and 3 even selected ``never''.}
26 participants agreed that “VPA platform should provide a policy compliance training for developers before they develop skills” (Q3) and 24 participants thought “VPA platform should check the skill code about policy violation” (Q4). 
\song{
Although 17 developers selected they would “always” or “most of the time” check whether privacy policy and code are consistent (Q5),
and 23 developers mentioned they would “always” or “most of the time” check their code consistency before submission (Q6), \ProjectName{} identified hundreds of skills with inconsistency issues in our skill code dataset (details in Section~\ref{evaluation}). 
At last,}
most participants thought that a static analysis tool for checking skill code could be ``extremely'' or ``very” useful for policy-compliant skill development.

\song{Table~\ref{userstudy:skills} lists the policy violations in our example skills as well as users' response results. \liao{For each skill, we asked developers, ``Do you think this skill violates any Amazon policy requirements? If so, which policy does it violate?''.}
}
21 (62\%) participants reported that the second skill (S2) contains a policy violation (\ie, collecting kids personal data) and gave the right reason.
Only 13 (38\%) developers were aware that ``asking for a positive rating'' is a policy violation (S3). This partially explains why there are considerable published skills in the skills store violating this policy~\cite{SkillDetective:2022}.
For the other three policy-violating skills that ``call emergency responder'', ``predict gender'' and ``have \liao{a disallowed} invocation name'' \liao{(invocation name has two words and one word is a definite article ``the'')}, only 7, 1 and 5 developers correctly reported the violations, respectively.
\song{
In addition, for the 3 skills without a violation, 21 developers wrongly thought they contained some policy violations. 
}
Our user study results show that there is a notable gap between VPA's policy requirements and skill developers' practices and perceptions.
Such observations motivate us to develop a static analysis tool to facilitate third-party developers in developing policy-compliant skills as well as improving the quality of their code.


\begin{table}[t]
	\centering
	\vspace{5pt}
	\resizebox{8.5cm}{!}{
		
			\begin{tabular}{  
			>{\centering\arraybackslash}m{1.2cm} >{\arraybackslash}m{2.4cm} >{\arraybackslash}m{3.7cm}
			>{\centering\arraybackslash}m{2.3cm}
			}
				\Xhline{5\arrayrulewidth}
				{\bf  Skill index}  
				& 
				{\bf Violation type}
				& 
				{\bf Skill output that contains a policy violation }
				& {\bf \# of users that reported a violation} 
				\\
				\hline
				S2 & Collect kids data & Welcome to cake walk for kids! Can you tell me your birthday?  & 21 (62\% \cmark) 
				\\
				\hline
				S3 & Ask for a positive rating & If you like this skill, please give us a 5 star rating. &13 (38\% \cmark) 
				\\
				\hline
				S6 & Call emergency\quad responder & We can also call hospital or emergency responders. & 7 (21\% \cmark) 
				\\
				\hline
				S7 & Predict gender & Would you like Alexa to predict the gender of your baby? & 1 (3\% \cmark)  
				\\
				\hline
				S8 & \liao{Disallowed} invocation name	 & the birthday & 5 (15\% \cmark)
				\\
				\hline
				S1, S4, S5 & No violation & N/A & 21 in total (21\% \xmark) 
				\\
				\Xhline{5\arrayrulewidth}
	\end{tabular}
	}
	\caption{Policy violations in example skills and the responses from 34 participants (\cmark~means a correct answer and \xmark~means an incorrect answer). 
 }
	\label{userstudy:skills}
	\vspace{-25pt}
\end{table}

\section{SkillScanner Overview}

\begin{figure*}[!h]
	\begin{center} 
		\includegraphics[width=0.9\textwidth]{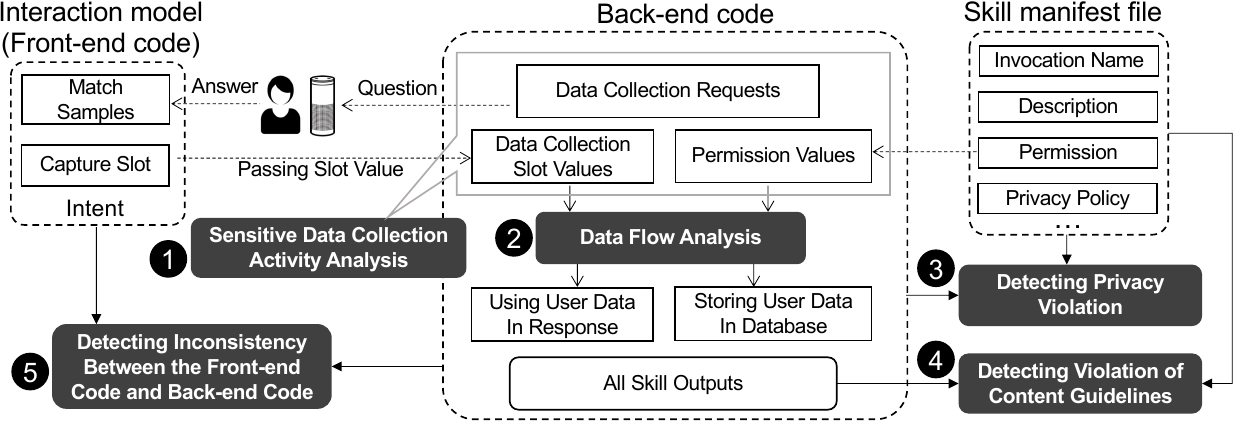}
	\end{center} 
	\vspace{-10pt}
	\caption{System Overview of \ProjectName{}.}
	\label{SystemOverview}
	\vspace{-10pt}
\end{figure*}

{\bf Threat Model.}
The proposed \ProjectName{} is mainly designed for benign third-party developers to identify potential policy violations in skills at the development phase. We assume that inexperienced developers may unknowingly develop and publish policy-breaking skills to the public, since our user study results in Section~\ref{userstudy} demonstrate that they are not completely aware of the related policies. These problematic skills could pose privacy, safety, and security threats to VPA users. 
\ProjectName{} is designed to run locally (which also protects developers' proprietary code) to identify potential policy violations and inconsistencies in skill code. Thus, we assume that the source code of skills is provided by developers as inputs of the tool. 
\liao{Since \ProjectName{} aims to help benign developers improve the policy compliance of their skills, for adversarial application scenarios such as a developer intends to violate the policy, it is out of this work's scope. \ProjectName{} can be potentially used by the Amazon Alexa platform for policy violation detection if developers provide access to their skill code to Amazon Alexa during the skill certification process.
}




{\bf System Overview.} With the goal of ensuring privacy and policy compliance in the Amazon Alexa ecosystem,
\ProjectName{} mainly focuses on detecting violations of privacy requirements~\cite{PrivacyRequirements} and skill content guidelines~\cite{PolicyTesting}
\song{(which all skills must adhere to according to the Amazon Alexa documentation),}
as well as inconsistencies in skill code. 
Figure~\ref{SystemOverview} shows the design overview of \ProjectName{}. 
To detect data collection related issues, we first extract data collection activities by analyzing both the back-end code and front-end code~(\ding{182}). 
A skill can collect user data 
through the conversational interface or using \song{Alexa's built-in} permission APIs. 
For collecting user data through the conversational channel, a skill normally asks questions such as ``How old are you?''
in the back-end code. 
We extract all potential skill outputs 
and then conduct an NLP analysis of these outputs to check whether 
they collect any user data. 
Then we find corresponding data collection slots in the front-end code by checking whether the slot name and slot type contain any data collection keywords. 
Developers can also directly collect user data by requesting for specific permissions to access user data in the back-end code. We identify such data collection activities by checking if any permission API is used in the code.
Next, \ProjectName{} tracks how the collected data is used in skill code by conducting a taint analysis (\ding{183}).
The source of taint analysis can be a data collection slot value or the returned value from a permission API in the back-end code. A skill may use the collected user data in response (\eg, to establish a rapport with users), store the data in a local database, or never use the data in the back-end code.

After the taint analysis, 
we check whether the data collection and data usage in code are consistent with what developers claimed in the skill manifest file
(\ding{184}).
For example, developers are required to provide a privacy policy document to outline data collection and usage, and there might be inconsistencies between the privacy policy and the actual data collection activities in the code.  
In addition, developers may request for more or less permissions than they actually need. 
Amazon also defines a diverse range of content guidelines, ranging from skill output content safety to user review manipulation (\eg, explicitly requesting 5-star ratings). 
To ensure compliance with content guidelines, \ProjectName{} analyzes all skill outputs to check whether there exists any violation of the content guideline (\ding{185}). 
At last, a poorly-written interaction model (\ie, the front-end code) may trigger incorrect intent and return users with content they don't expect. Such poor user experience can lead to user frustration and a decrease in user engagement. 
To facilitate developers to improve the quality of
their code, \ProjectName{} also detects inconsistencies in the back-end and front-end code (\ding{186}). With the help of \ProjectName{}, developers can improve their code quality based on the policy violation report and provide a better user experience of skills.

\section{System Design and Implementation} 

\subsection{Sensitive Data Collection Activity Analysis}\label{CollectionActivity}
To learn what data a skill is able to collect, we need to analyze both the back-end and front-end code. The front-end code is stored in a JSON file and the back-end code is written in JavaScript or Python language.
We first define the sensitive/personal data types that are considered in \ProjectName{}, including 1) 24 types of PII (personally identifiable information) from a NIST (National Institute of Standards and Technology) report~\cite{mccallister2010guide} \song{and Amazon~\cite{amazon_pii}}, 2) Amazon's built-in slot types~\cite{AlexaBuiltinSlot} such as ``Amazon.Person'' for getting the user name and ``Amazon.US\_City'' for US cities; 3) specific types
of data that can be collected through the permission APIs~\cite{AlexaPermission}; and 4) common health information.
The keywords about data collection are listed in Table~\ref{table:datacollection}. 




\subsubsection{{Identifying conversational sensitive data collection activities}}
\label{data_collection_activity}

Our analysis starts from identifying data collection requests (\eg, ``What is your name'' or ``Please tell me your name'') from a skill's possible outputs.
We search for all strings in the back-end code (excluding comments).  
Given an output extracted from a string, we apply an NLP-based method and use the Spacy library~\cite{spacy} to check if any data collection keyword is used as a noun because some keywords
like ``address'' 
can be used as a verb. 
Developers may also provide their own data,
so we check whether any PII data is used as a noun following the word ``your''. If a sentence contains ``your + sensitive data collection noun'', we consider it as a data collection request. In addition, we check the output with
a list of common sentences of personal data collection~\cite{SkillDetective:2022}, such as ``how old are you'' or ``what can I call you'' to improve accuracy.
To track whether user provided data can be properly processed and passed to the front-end code, we need to analyze slots and sample utterances of each intent in the front-end code. 
As introduced in Section~\ref{background}, {when users hear a question and make a reply, there should be an intent and slot for processing the reply.}
If a slot includes any data collection keyword listed in Table~\ref{table:datacollection} in its name or slot type, 
we define such slot as a sensitive data collection slot, \eg, a slot named as ``username'' with the slot type ``Amazon.FirstName''.
Similar to identifying sensitive data collection requests, if a sample utterance includes ``my + \{data collection slot\}'', we consider it as a data collection utterance.

We capture conversational data collection activities of a skill by checking the following three conditions: 1) whether there exists a data collection request asking for user data;
2) whether there is any sample utterance that can be matched to potential user replies to this data collection request; and 3) whether a slot of the sample utterance is used to capture sensitive user data.
The back-end code could not correctly get user data through the conversational interface if missing any one of these conditions.
Once we identify conversational data collection activities in the skill code, 
the corresponding slot values are used as taint sources in our data flow analysis in Section~\ref{TaintSources}.

\begin{table}[t]
	\centering
	\resizebox{8.5cm}{!}{
		\begin{tabular}{  >{\centering\arraybackslash}m{2.5cm} >{\centering\arraybackslash}m{7.4cm} }
		    \toprule[1.5pt]
			{\bf Personally Identifiable Information (PII)~\cite{mccallister2010guide,amazon_pii}} & Address, Name, Email, Birthday, Age, Gender, Account, Location, Contact, Phonebook, Profession, Income, Zipcode, Postal code, Phone number, Passport number, Driver license number, Bank account number, Debit card number, Credit card number, Credit card verification code, Taxpayer identification number,  
			Social Security number (SSN),
			Vehicle identification number (VIN)
			\\
			\bottomrule[1pt]
			{\bf Amazon's built-in slot types} & FirstName, Person, US\_FirstName, PhoneNumber, PostalAddress, Region, RelativePosition, City, US\_City, AdministrativeArea, StreetAddress, StreetName, US\_State, Professional, ProfessionalType 
            \\
			\bottomrule[1pt]
			{\bf Data types supported by permission APIs} & 
            Profile: Name, Given\_name, Email, Mobile\_number
            Devices: Address, CountryAndPostalCode, Geolocation
			\\
			\bottomrule[1pt]
			{\bf Health Information} & Height, Weight, Blood group, Blood pressure, Blood glucose, Blood Oxygen, Heart rate, Body temperature, Sleep data, Fat percentage,
            Mass index, Waist circumference, Menstruation, Period
            \\
			\bottomrule[1.5pt]
		\end{tabular} 
	}
	\caption{Keywords related to personal data collection.}
	\label{table:datacollection}
	\vspace{-25pt}
\end{table}

\subsubsection{Identifying data collection permission requests}

Another way for a skill to collect user data is to request permissions for customer information, which is from the user's Alexa account.
Amazon states that ``Alexa skills might require personal information from the customer in order to provide relevant information in skill responses or to complete transactions''. Table~\ref{table:datacollection} lists the data types that are considered for permission-based data collection activity analysis in~\ProjectName{}.
In this method, developers first request for permissions in the Amazon Alexa developer console, and permission request information is stored in the skill manifest file. Before accessing data, a skill needs to get a user's grant to access permission data when a user first enables the skill.
Once granted, the skill can retrieve the requested permission data using different methods in the back-end code: either directly {using certain pre-defined endpoints or calling permission APIs}. Since they exhibit distinct patterns, we can easily identify data collection permission requests in the back-end code. {For example, in line 18 of Listing~\ref{lst:code1}, the user email data is obtained from a permission request, and thus we take this value as the taint source of the data flow analysis.}






\subsection{Data Flow Analysis}\label{TaintAnalysis}

If a skill contains any data collection activity, we track how the collected user data is used by conducting a taint analysis in the back-end code. Our objective is to check whether user data is properly used or saved into databases, which is useful for detecting privacy related violations in Section~\ref{PrivacyDetection}. For example, in Listing~\ref{lst:code1}, we know the slot value \texttt{userName} and permission value \texttt{profileEmail} (in lines 8 and 18) are data collected from users. By tracking their data flows, we learn that these two values are used for generating customized responses (in lines 12 and 20) and storing to the database (in line 10).

\vspace{1pt}
\begin{lstlisting}[caption={Real-world skill code with  sensitive data collection.},label={lst:code1}, style=myStyle, frame=single, language=C, commentstyle=\color{mGreen},
] 
const LaunchRequestHandler = {
  handle(handlerInput){
    const speakOutput = 'Hello! What is your name?'; // Ask for user name
    return handlerInput.responseBuilder.speak(speakOutput).reprompt(speakOutput).getResponse(); 
},};
const CaptureUserNameHandler = {
  handle(handlerInput){
    const userName = handlerInput.requestEnvelope.request.intent.slots.name.value;          // Get name from slot
    const sessionAttributes = handlerInput.attributesManager.getSessionAttributes();
    sessionAttributes.name = userName; // Store name in database
    const speakOutput = 'Thanks ${userName}';
    return handlerInput.responseBuilder.speak(speakOutput).reprompt(speakOutput).getResponse();       // Use name for response
},};
const EmailIntentHandler = {
  handle(handlerInput){
    const { serviceClientFactory, responseBuilder } = handlerInput;
    const upsServiceClient = serviceClientFactory.getUpsServiceClient();
    const profileEmail = await upsServiceClient.getProfileEmail();  // Get email from permission
    const speechResponse='Your email is ${profileEmail}';
    return handlerInput.responseBuilder.speak(speechResponse).reprompt(speechResponse).getResponse();  // Use email for response
},};

\end{lstlisting}
\vspace{-10pt}

\subsubsection{Taint sources}\label{TaintSources}
A skill can obtain user data by retrieving slot values or permission values, which are the sources of our taint analysis. We describe how \ProjectName{} locates taint sources in the back-end code.

{\bf Slot values as taint sources.} There are two ways in a skill's back-end code to retrieve slot values. 1) The first method is to send requests and get slot values. 
The back-end code can use a request handler to handle intent and get slot values with ``handlerInput.requestEnvelope.request.intent.slots''. 
2) The second method is using the Alexa Skills Kit (ASK) SDK, which provides several functions for retrieving slot values.
Developers can directly call ``Alexa.getSlotValue'' 
and use slot name as parameter to get a slot value.
Accordingly, after identifying conversational data collection
activities in a skill \song{(in Section~\ref{data_collection_activity})}, we use the ``intent.slots'' and ``Alexa.getSlotValue'' as sources in our taint analysis.


{\bf Permission values as taint sources.} 
A skill obtains permission values either through pre-defined 
endpoints or permission APIs in the back-end code.
1) A skill can directly access pre-defined endpoints to get customers' contact information
and setting information such as address or postal code. 
Table~\ref{table:endpoints} lists the pre-defined endpoints in Alexa, including two types of user data: device address (denoted as ``/v1/'') and customer profile (denoted as ``/v2/'').
For the customer profile, Alexa provides information about the name (full name and give name), email and mobile phone number. 
\ProjectName{} searches for all string values in code with endpoints listed in  Table~\ref{table:endpoints} and treats the returned values of these endpoints as the taint sources. 2) 
The second approach of obtaining permission values is to use Alexa Service APIs. Amazon provides ``DeviceAddressServiceClient'' for getting device addresses and ``UpsServiceClient'' for customer profiles. For example, a skill can get the device address through 
``DeviceAddressServiceClient.getFullAddress()''
or obtain user email with 
``UpsServiceClient.getProfileEmail()'' (as shown in line 18 of Listing~\ref{lst:code1}). We also take the returned values of these permission requests as the taint sources.

\begin{table}[t]
	\centering
	\resizebox{8.0cm}{!}{
		\begin{tabular}
		{  >{\arraybackslash}m{3.4cm} >{\arraybackslash}m{5.8cm} }
		\toprule[1.5pt]
		{\bf Requested Information} & \bf{Endpoints} \\
		
		\bottomrule[1.0pt]
        Device Address/ Device Country and Postal Code& 
        \url{/v1/devices/{deviceId}/settings/address}\quad (/countryAndPostalCode)\\

        \bottomrule[1.0pt]
		Full Name/ Given Name/ Email/ Phone Number & \url{/v2/accounts/~ current/settings/Profile.name}  (/givenName/email/mobileNumber) \\
		
		\bottomrule[1.0pt]
        Full Name/ Given Name/ Phone Number	&
        \url{/v2/persons/\textasciitilde current/profile/name}\qquad (/givenName/mobileNumber)\\

			\bottomrule[1.5pt]
		\end{tabular} 
	}
	\caption{Pre-defined endpoints to obtain permission data}\label{table:endpoints}
	\vspace{-25pt}
\end{table}


\subsubsection{Taint sinks}
We mainly track two data usage cases in a skill, 1) using the collected user data in the response or 2) storing the data in a local database.
Typically, a skill uses ``handlerInput.reponseBuilder'' to generate a verbal response by calling ``.ask(output)'', ``.speak(out-put)'', or ``.reprompt(output)''. To track if  
the collected data is used in generating customized
skill responses, we search for these functions in the back-end code and use them as taint sinks.
Skills can also save the user data to a database so that they can use them for other purposes. If a skill needs to store data in a database, it usually first builds the database connection. 
{Developers need to call specific APIs to create or retrieve data from a DynamoDB database provided by Amazon, or save data to the Alexa session attribute.}
These database APIs are considered as taint sinks in \ProjectName{}. 

\begin{figure}[!h]
	\begin{center} 
		\includegraphics[width=0.45\textwidth]{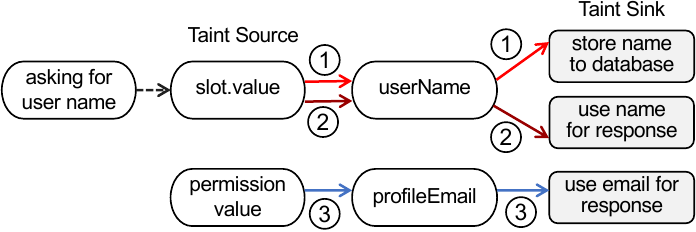}
	\end{center} \vspace{-10pt}
	\caption{Data flows corresponding to Listing~\ref{code1}.}
	\label{code1_taintanalysis}
	\vspace{-10pt}
\end{figure}

\subsubsection{Tracking data flows}


Since a skill can be written in different programming languages, we use the CodeQL tool~\cite{codeql} that supports multiple programming languages to track data flows of different variables in skill code. \liao{We demonstrate the performance of CodeQL in Section~\ref{DataFlowAnalysis} and it performs well in our results.} Figure~\ref{code1_taintanalysis} shows the data flow analysis results corresponding to the code in Listing~\ref{lst:code1}. \ProjectName{} identifies three data flows, the first one is from the sensitive data collection slot value (\texttt{userName} in line 8 of Listing~\ref{lst:code1}) to the database usage (line 10 of Listing~\ref{lst:code1}) and the second data flow goes to a speaker output (line 12 of Listing~\ref{lst:code1}). The third data flow is from the permission request of user email address (\texttt{profileEmail} line 18 of Listing~\ref{lst:code1}) to a speaker output (line 20 of Listing~\ref{lst:code1}).





\subsection{\song{Detecting Privacy Violation}}
\label{PrivacyDetection}
Among the various policy violation problems reported by recent research~\cite{SkillExplorer:2020,Shezan:VerHealth:2020,Lentzsch:NDSS:2021, SkillDetective:2022, Jide:2022:WWW, Song:2020:ACSAC}, many issues are related to the privacy policy. In particular, Jide~\etal~\cite{Jide:2022:WWW} performed a longitudinal measurement of privacy policies of skills across three years. They show that many developers were not engaging in good data practices, which is still an unresolved issue.
In the skill manifest file, we can obtain several types of skill information such as skill name, category, description, privacy policy, permission, etc. \ProjectName{} detects problematic privacy policies and inconsistencies between the back-end code and the disclosed information provided by developers according to the skill manifest file.



\subsubsection{Problematic privacy policy}
\begin{figure}[!h]
	\begin{center} 
        \subfloat [Missing a privacy policy (from skill ``Boba Maker'')]
        {
        \includegraphics[width=0.4\textwidth]{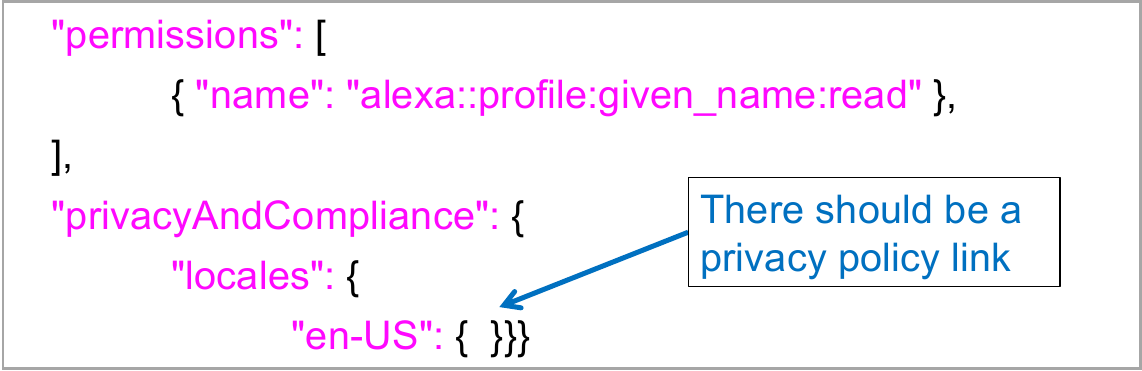}
        \label{subfig:lackprivacypolicy}
        }
	\end{center} 
	\begin{center} 
	\subfloat [Incomplete privacy policy (from skill ``feed me now'')]
	    {
		\includegraphics[width=0.4\textwidth]{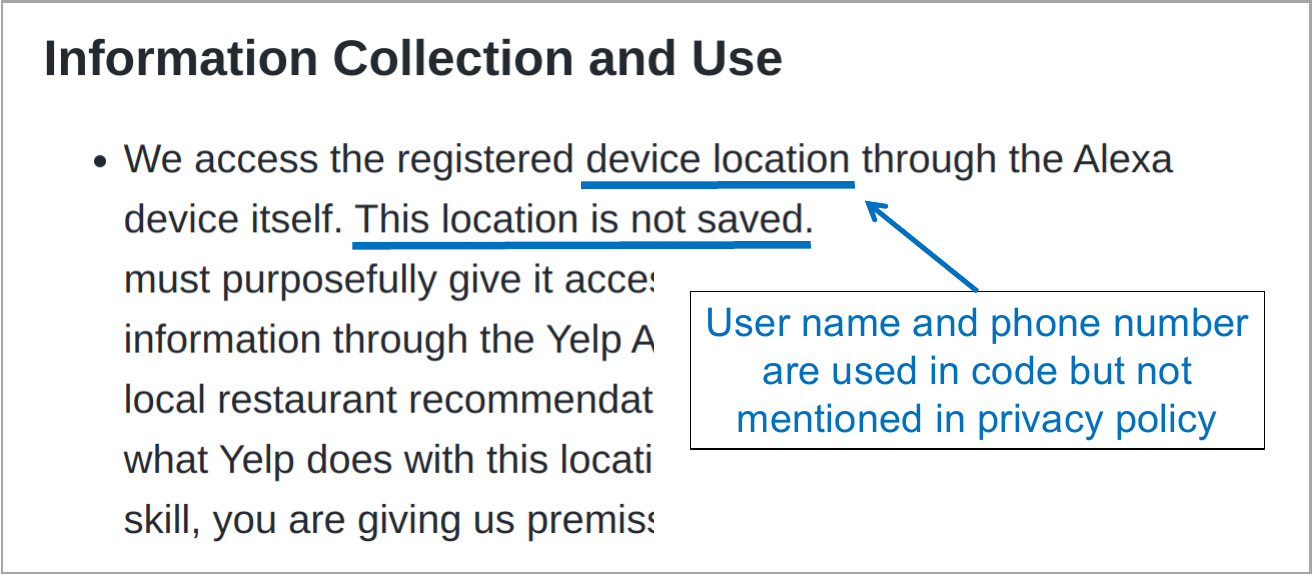}
		 \label{subfig:incompleteprivacypolicy}
		}
	\end{center} 	
	\vspace{-10pt}
	\caption{\song{Examples of problematic privacy policies.}}
	\label{privacypolicy}
	\vspace{-10pt}
\end{figure}

According to Alexa's privacy requirements~\cite{PrivacyRequirements}, it requires that skills with data collection activities must provide a privacy policy URL that links to a legally adequate privacy policy webpage. The privacy policy URL will be displayed to end users on a skill's web page in the skills store. 
Given a skill source code package, \ProjectName{} first checks if the skill provides a valid privacy policy link (\ie, not a broken link or leading to an unrelated webpage). 
Based on the results of the sensitive data collection activity analysis in Section~\ref{CollectionActivity} and taint analysis in Section~\ref{TaintAnalysis}, we check whether a skill's data collection and usage are properly disclosed in its privacy policy.
{If a privacy policy is provided but the data practices are not mentioned in it, such privacy policy is incomplete since it doesn't disclose the sufficient information.}
\song{Figure~\ref{privacypolicy} shows two skills that miss a privacy policy or provide an incomplete privacy policy.}
{For each sentence in a privacy policy, we check whether it mentions any sensitive data collection or data storage activity. For the data collection activity, we check whether a data collection verb, such as ``collect'', ``ask'' or ``access'' and the collected data appear in the same sentence. For the data storage activity, we check if the ``store'' verbs, such as ``keep'', ``retain'' or ``store'', and collected data are in the same sentence~\cite{yu2016can, breaux2014eddy}.}

\subsubsection{Over-privileged data \song{collection permission} requests}
Over-privileged data collection permission requests refer to a skill requesting more permissions than it requires~\cite{su2020you}, and thus the skill violates the principle of least privilege. 
With the taint analysis, \ProjectName{} checks if a skill requests for permission data but never uses it in the back-end code. \ProjectName{} also detects if a skill \song{tries to use a permission to retrieve user data but doesn't ask for that permission first}.


\subsubsection{Inconsistent data collection disclosure to Alexa platform}
Amazon provides a selection for developers before submitting a skill for certification that asks ``Does this Alexa skill collect users' personal information?'' and developers can select ``yes'' or ``no'' \song{(they can also choose to not answer this question)}. 
{Different from a privacy policy, which is 
provided for end users, such a question is for developers to disclose data collection behavior to the Alexa platform and certification team.}
Researchers in~\cite{Long:2020:CCS} observed that if a developer 
selects no to this question but his/her skill collects data through the conversational interface, it is more likely the skill could bypass the certification process.
The developer's answer information can be found in the skill manifest file. 
Based on the results of our data collection activity analysis, \ProjectName{} checks if any inconsistent data collection disclosure between the back-end code and developers' answers.



\subsection{Detecting Violation of Content Guidelines}
The Amazon Alexa platform has defined a list of content guidelines~\cite{PolicyTesting}, describing the inappropriate content that skill should not deliver to users. 
These involve disturbing content, false information, profanity, etc. 
In this section, we focus on
ensuring the content safety and compliance of invocation names.


\subsubsection{Content safety violations}\label{contentsafety}
\ProjectName{} mainly checks three types of content violations: toxic content 
, user review manipulation 
and content safety of external websites involved in a skill's back-end code.
\ProjectName{} detects whether there exists potential toxic content in skill outputs by using Google's Perspective tool~\cite{Perspective}, which is a machine learning based tool for detecting toxic content.
\ProjectName{} extracts all possible skill outputs 
from a skill's back-end code. If any skill output's \textit{Toxicity} and \textit{Profanity} scores are higher than the threshold (0.9), \ProjectName{} generates a warning of possible toxic content for developers. 

To avoid user review manipulation, Amazon Alexa enforces a policy prohibiting ``Explicitly requests that users leave a positive rating of the skill''. \ProjectName{} checks whether a skill explicitly asks for the user to provide a positive rating in the skill outputs as well as the skill description.
We detect this violation using an NLP-based method by checking whether keywords such as ``5 star'' or ``five star'' are used followed by verbs ``give'' or ``leave''.


A skill's back-end code may contain external websites from which the skill can call HTTP requests to get data (\eg, pre-recorded audio streams and images) from an external data source. Modern VPA devices (\eg, Amazon Echo Show) can display different media files including text, images and even movies.
We consider three different cases that websites are involved in skill code: 1) retrieving audio data such as ``mp3'' files; 2) retrieving image data such as ``jpg'' files; or 3) retrieving textual content from external websites.  
For audio and image files, we first download them and then use speech-to-text~\cite{SpeechRecognition} and image-to-text tools~\cite{pytesseract} to extract the textual data from these files. The output data is treated as normal skill output and checked whether it includes any policy violation such as data collection or toxic content. 
We also detect if an outside website is flagged by ViruslTotal~\cite{virustotal} as a suspicious or malicious website. For the skills retrieving data from external websites, we check whether the webpage contains any inappropriate content according to Amazon Alexa's content guidelines.

\subsubsection{Non-compliance of invocation names}
Under the Amazon Alexa’s content guidelines, there is a special category for skill invocation names and several requirements are listed, such as ``one-word invocation names are not allowed'' and ``the invocation name must not contain any of the Alexa skill launch phrases''. 
More details about invocation name requirements and disallowed words as well as representative violation cases we found using \ProjectName{} are listed in Table~\ref{table:invocation_name}.
Since a skill's invocation name information is stored in the front-end code, \ProjectName{} extracts a skill's invocation name and checks whether it violates the policy requirements.

\begin{table}[h!]
	\centering
	\vspace{0pt}
	\resizebox{8.4cm}{!}{
		\begin{tabular}
		{  >{\arraybackslash}m{7cm}
		    >{\centering\arraybackslash}m{3cm}
		}
		
		\toprule[1.5pt]
		\bf{Requirements on Invocation Name} & \bf{\shortstack{Example invocation \\names with violation}}	\\
		\bottomrule[1.0pt]
		 One-word invocation names are not allowed, unless the invocation name is unique to your brand/intellectual property. & \shortstack{beeper, \\jokes}
		 \\
        \bottomrule[1.0pt]
        Two-word invocation names are not allowed if one of the words is a definite article (``the''), indefinite article or preposition.
        & the template, \qquad\qquad the radiator\\
        \bottomrule[1.0pt]
        The invocation name must not contain any of the Alexa skill launch phrases (such as ``play'', ``launch'', ``open''), connecting words (such as ``to'', ``about'', ``and'') or wake words (such as ``Alexa'', ``skill'', ``app'').
        & \shortstack{play radio,\\ to jeff,\\ video app}\\
        \bottomrule[1.0pt]
        The invocation name must contain only lower-case alphabetic characters, spaces between words, and possessive apostrophes. 
        & AlTranslate,\quad\quad Ryan's note\\
        \hline
        \song{The invocation name should be distinctive to ensure users can enable your skill.} & \shortstack{hello world\\ (used by 85 skills)}\\
			\bottomrule[1.5pt]
		\end{tabular} 
	}
	\caption{Invocation name requirements~\cite{PolicyTesting}.}\label{table:invocation_name}
	\vspace{-25pt}
\end{table}

\subsubsection{Category-specific violations}
Amazon defines specific policies for selected categories, such as ``Kids'' and ``Health \& Fitness'' categories.
For skills in the Kids category, they can't include unsuitable content, direct users to 
external websites, or collect any personal information~\cite{KidsPolicy}, regardless of whether a privacy policy is provided.
For data collection in Kids skills, we used the same method as described in Section~\ref{CollectionActivity} to detect \song{data collection requests} and the method in Section~\ref{contentsafety} to check skill outputs about external websites and toxic content. 
For the Health category, Alexa requires that the skill must ``include a disclaimer in the skill description stating that the skill is not a substitute for professional medical advice''~\cite{HealthPolicy}. We checked whether the words ``medical advice'', ``educational purpose'' or ``information purpose'', which are keywords taken from \song{an example disclaimer provided by Alexa}, show up in a skill's description. 

\hc{similarly to 4.3, in 4.4 we may also need to be explicit on how we achieved the detection in a principled approach, and how the static analysis results are used; our overview figure shows the connection between the taint analysis and detection steps, we should be consistent with that overall design}

\subsection{Detecting Inconsistency Between the Front-end Code and Back-end Code}


\song{As mentioned in Section~\ref{CodeStructure}, intents (as well as sample utterances and slots) in a skill's front-end code and the back-end code are tightly coupled.}
Typically, developers define a \song{data collection request} in the back-end code (\eg, what's your name), create an intent based on possible user replies in the front-end code, and then create the corresponding intent handler for processing user replies in the back-end code.
Inconsistency among them could trigger incorrect intents and impact user experience. For example, if a skill asks for user data through the conversation but doesn't have an intent for processing the user reply, the skill will trigger a built-in intent ``FallbackIntent'' which replies ``Sorry, I don't know that. Please try again''.
\liao{Amazon requires developers ``reviewing the intent schema, the set of sample utterances, and the list of values for any custom slot types developers have defined to ensure that they are correct, complete, and adhere to voice design best practices''~\cite{AlexaTesting}. So,
\ProjectName{} also} detects potential bugs {and improves the code quality} early in the skill development lifecycle \song{because of its influence on skill function and user experience.}



\ProjectName{} detects three cases of code inconsistency. 1) 
A skill's back-end code has a \songl{request} asking for user data, but there is no slot in the front-end code for processing the user reply. 2) A skill defines a slot to obtain users' possible data but there is no \songl{request} in the back-end code asking for user data. 3) {A data collection slot or an intent} doesn't have any sample utterances. If so, they could not match with any user reply, and thus the slot or intent will never be triggered.
The second case is actually a code vulnerability~\cite{Long:2020:CCS, su2020you} due to the fact that 
the Amazon Alexa platform does not require a re-certification when a change is made in the back-end code of a skill.
If a skill has a data collection slot but doesn't have a \song{data collection request} asking for user data, after the skill has been certified and published in the skills store, developers can arbitrarily change the back-end code and ask for any type of user data. 
\liao{Benign developers may collect user data for non-malicious purposes but unintentionally violate policy after the code update (\eg, collecting user names and using them to establish a rapport with users). \ProjectName{} reminds developers of such potential policy violations.}
Since Amazon Alexa can not check these consistencies at the skill certification phase, \ProjectName{} facilitates developers to detect potential code defects that are difficult to find for improvements. 


\section{Evaluation}
\label{evaluation}
\hc{we may describe a bit about our experiment setup such as datasets used (e.g., the VPA platform and the skills evaluated)}

In this section, we evaluate \ProjectName{} by answering the following three research questions:

\textbf{RQ1}: {\em How effective is \ProjectName{} in capturing sensitive data collection and usage behaviors in skills?  ($\S$~\ref{DataCollectionAnalysisResult} and $\S$~\ref{DataFlowAnalysis})}

\textbf{RQ2}: {\em How effective is \ProjectName{} in identifying policy violations in open source skills?  ($\S$~\ref{PrivacyViolation}, $\S$~\ref{ContentViolation}, and $\S$~\ref{CodeInconsistency})}

\song{\textbf{RQ3}: {\em \ProjectName{} found many similar violations in different skills. What's the main reason for these similar policy violations? ($\S$~\ref{publishedskills})}}

	
	

\subsection{Skill Source Code Collection \& Setup}
\song{Before discussing the evaluation results, we explain how we collected open source skill code from GitHub and the setup that we used to perform the evaluation.}

In contrast to traditional apps on smartphone platforms (\eg, Android or iOS) where apps run on host smartphones,
a skill's back-end code runs on the developer's server, and is not available even for the Amazon Alexa platform. We can not obtain skill code from the Alexa's skills store. Instead, we collected open-source skill code from GitHub, which is one of the largest open-source project platforms. The unique file structure 
\song{(\eg, every skill comes with a front-end code file ``en-US.json'')}
of skill packages allows us to accurately locate skill code in GitHub. Specifically, we searched a combination of keywords ``en-US.json'' (front-end code filename), ``interactionModel'', ``languageModel'' and ``intents'' for the front-end code file using GitHub \liao{search} API.
Finally, we were able to find 2,451 skill repositories \song{after removing duplicate skills} and we shared the dataset to facilitate future research (\url{https://github.com/CUSecLab/SkillScanner}). Table~\ref{table:skilldatanumber} shows the statistics of skill code in our dataset.

\begin{table}[h]
	\centering
	\resizebox{8.0cm}{!}{
		{ 
			\begin{tabular}{
			>{\centering\arraybackslash}m{2.5cm} >{\centering\arraybackslash}m{4.5cm}
			>{\centering\arraybackslash}m{1.5cm}
					}
				\Xhline{3\arrayrulewidth}
				\multirow{3}{*}{Front-end code} & Total \# of custom intents & 7,880 \\
				\cline{2-3}
				 & Total \# of slots & 5,702  \\
				\cline{2-3}
				& Total \# of sample utterances & 67,884  \\
				\hline
				{Back-end code} & Total \# of functions & 26,216 \\
				\hline
				\rowcolor[gray]{.9}
				&Total \# of skills &
				2,451 \\
				\Xhline{3\arrayrulewidth}
			\end{tabular}
		}
	}
	\caption{Statistics of skill code in our dataset.}
	\label{table:skilldatanumber}
	\vspace{-20pt}
\end{table}

Since the skills in our dataset were written either in Python or Node.js, our data flow analysis module has two versions.
We ran \ProjectName{} on a server with \songl{3.0 GHz 6-core CPU and 16GB RAM}. 
Finally, \ProjectName{} analyzed 2,451 skills in total and obtained 797,501 skill outputs, 1,656 HTML webpages, 1,439 mp3 files and 932 images in our experiments. 
\song{The output of \ProjectName{} is a report that notifies developers about potential policy violations or bugs and the corresponding code location. Then, developers can fix these issues, otherwise the skill may fail in the certification process.}

\song{By comparing skills' names, developers and descriptions of skills in our dataset and skills in the Alexa skills store, we found that 93 skills in our dataset have been published in the skills store.
We did notice that there exists ``toy'' projects in our dataset, which have a single contributor, few commits, and non-informative readme files. 
As previous works demonstrate, most skills are developed by third-party developers and these skills are easy to be published~\cite{Long:2020:CCS}, and also the quality of these published skills is not good~\cite{SkillExplorer:2020, SkillDetective:2022}. 
In our dataset, 15\% of skills that are published have a repository with 1 contributor, less than 10 commits and lack a readme file. Since these low-quality skills are also possible to be published to the skills store, we didn't remove them from our dataset.
}

\subsection{Sensitive Data Collection Activity Analysis}\label{DataCollectionAnalysisResult}
\vspace{5pt}
After getting all possible outputs from each skill's back-end code, we observed that 321 skills contain sensitive data collection requests
asking for user data through the conversational interface.
{197 of them have sensitive data collection slots in front-end code to process the collected data. For the other 124 skills that don't have a data collection slot, we will discuss this issue in \liao{Section~\ref{lackanintent}.}} 
Over half of the skills ask for the name information followed by the birthday, email and address information. Skills also ask for seven other types of data such as age, gender, location, phone number or zip code. The most common data collection requests are ``You can introduce yourself by telling me your name'' and ``What is your name?''. 
For example, the skill ``E-Nurse'' speaks out ``Kindly tell me your name'' when it is invoked. 
{In the front-end code, it defines an intent called ``NameIntent'' and one slot ``name'' with slot type ``Amazon.FirstName''. Two sample utterances are ``my name is \{name\}'' and ``I am \{name\}''.}
In addition, we found that 133 skills request for permission data. For example, a skill named ``Pizza Search''
requests for 7 permissions including five sensitive data collection permissions for address, location, email, mobile number and name information.
Overall, \ProjectName{} achieves high accuracy for the sensitive data collection activity analysis.
After manually checking all the results, only 15 sentences are false positive cases out of 767 sensitive data collection requests, with an accuracy of 98\%. 
\ProjectName{} achieves 99.7\% accuracy in identifying data collection slots and 100\% for permission requests.
\songl{Most false positives occur in sentences that include ``your + sensitive data collection noun'' but do not ask for data, such as ``Your location is around...''.
The recall rates for each part are \ls{84\%, 93\%, and 91\%}, respectively. The false negatives 
are because of the diverse ways to ask for user data (\eg, ``How should I call you?'' and ``Please specify an address.'')
or inability to locate and read the skill file.}



\subsection{Data Flow Analysis}\label{DataFlowAnalysis}

\vspace{0pt}
\begin{lstlisting}[caption={An example skill that requests the email and user name but doesn't fully use the requested data and lacks a privacy policy.},label={lst:code3}, style=myStyle, frame=single, language=C, commentstyle=\color{mGreen},
] 
//back-end code:
const ScheduleTripIntentHandler = {
    ...
    const email = await upsServiceClient.getProfileEmail();                       // Get user email
    const givenName = await upsServiceClient.getProfile GivenName();         // Get user name
    ...
    const speechText = 'Enjoy your trip, ${givenName}!';    // Only name is used, thus here is an over-privilege issue in the skill.
    return handlerInput.responseBuilder.speak(speechText).reprompt(speechText).getResponse();
};

//manifest file:
"permissions": [
  {"name": "alexa::profile:given_name:read"},
  {"name": "alexa::profile:email:read"
],
"privacyAndCompliance": {
  "locales": {
    "en-US": {}     // Here should be a privacy policy link but it is missing
  }},

\end{lstlisting}

For the 197 skills containing slots for sensitive data collection and 133 skills asking for at least one permission, we conducted a taint analysis of these sensitive data sources. {After manually checking the results,} \ProjectName{} achieves an accuracy of 95\% for slot data flow analysis and 89\% for permission data flow analysis.
The failure of data flow analysis often results from overlooked sinks that \ProjectName{} doesn't consider, such as using a user's name to make an appointment or using their location to find the nearest restaurant.
{For the 187 skills (after removing false positives from 197 skills) that contain slots for sensitive data collection,} we found that 8 skills don't use the data, 121 skills use the collected data in their outputs/responses and 91 skills save data to databases (one skill may contain multiple data usages). 
For the 118 skills asking for permission data, 42 skills don't use at least one permission data, which are over-privilege skills. 57 skills store data in databases and 30 skills use data for outputs or responses. 
Listing~\ref{lst:code3} shows an example that doesn't fully use the data it asks for. The skill requests the email and user name information but only the name is used in the skill response, and thus it is flagged as an over-privileged skill. \ProjectName{} can effectively detect such issues and warn developers to fix them during the development phase.


\subsection{Privacy Violation Detection}\label{PrivacyViolation}
When skills collect data from users, they should also provide complete and informative privacy policies for disclosing such data collection activities. Amazon Alexa also asks developers ``Does this Alexa skill collect users' personal information?'' during the skill submission phrase. However, we observed that many skills failed in disclosing their data practices in their privacy policy documents.


\vspace{-0pt}
\begin{table}[h]
	\centering
	\resizebox{8.5cm}{!}{
		{ 
			\begin{tabular}{
			|>{\centering\arraybackslash}m{2.5cm}|
			>{\centering\arraybackslash}m{1cm}|
			>{\centering\arraybackslash}m{2.3cm} |
			>{\centering\arraybackslash}m{3.4cm}| 
					}
				\Xhline{3\arrayrulewidth}
				\rowcolor[gray]{.9}  & \# of skills  & \# of skills missing a privacy policy & \# of skills  having an incomplete privacy policy \\
				\hline
				Data collection request & 321 & 240 (75\%) & 19 (6\%)\\
				\hline
				Ask for permission & 133 & 94 (71\%) & 23 (17\%) \\
				\hline
				Store slot value in database & 89 & 81 (91\%) & 4 (4\%) \\
				\hline
				Store permission value in database & 57 & 45 (79\%) & 6 (11\%) \\
				\Xhline{3\arrayrulewidth}
				
			\end{tabular}
		}
	}
	\caption{\song{Number of skills with privacy policy issues.}}
	\label{table:privacypolicyresults}
	\vspace{-25pt}
\end{table}

\subsubsection{Problematic privacy policy}
A privacy policy should clearly describe how user data is collected, used and shared. However, after checking the privacy policies of all data collection skills, 
we found that only 19\% of the skills provide a complete privacy policy. Table~\ref{table:privacypolicyresults} shows the breakdown of privacy policy issues in different types of data collection.
For the 321 skills with data collection requests, most of them have inconsistency issues between their data collection request and privacy policies. 75\% of skills don't provide a privacy policy and 6\% of skills provide an incomplete privacy policy that doesn't claim their actual data collection behavior.
For the 133 skills asking for data with permissions, 71\% of them lack a privacy policy and 17\% of them have an incomplete privacy policy. 

When it comes to how data is stored in databases, few skills mention that in privacy policies. For the 89 skills that store data collection slot value, 91\% of them lack a privacy policy and only 4 skills (4\%) mention the data would be stored. For the 57 skills that ask for permissions and store data in a database, only 21\% provide a privacy policy and 10\% all are useful. Other 6 skills provide broken or unrelated websites while one skill even says that it ``will not retain data'', which is deceptive.

Since the privacy policy link is extracted from the skill manifest file, a skill lacks a privacy policy if we couldn't find it. So for skills missing a privacy policy, \ProjectName{} achieves 100\% accuracy. For skills providing an incomplete privacy policy, we manually checked their privacy policy content. There are 3 false positive cases (already removed in Table~\ref{table:privacypolicyresults}) for data collection skills
and \ProjectName{} achieves an accuracy of 93\%.
\liao{The recall for detecting problematic privacy policy is \ls{86}\% and it's partially because of the false negatives in detecting data collection requests (details in Section~\ref{DataCollectionAnalysisResult}) or failure to read the manifest file to obtain the privacy policy link.}

\subsubsection{Over-privileged data requests}
The requested permission information can be extracted from the skill manifest file. 
To detect over-privileged permission data requests, we checked whether these permissions are used in the back-end code or not. {There are 42 skills that request for permissions but don't use them. 
There are also 18 skills that get permission data without requesting for permissions.} 

\subsubsection{Inconsistent data collection disclosure to Alexa platform}
Amazon asks developers whether their skills collect user data before submission and developers can select yes or no by themselves. The answers will be stored in the skill manifest file if a developer selects it. There are \song{126 skills with data collection in our dataset making a selection,
in which 99 skills} claim not using personal info while only 27 developers correctly select they use personal info. However, for the 27 skills, 4 skills don't provide a privacy policy and 16 provide an incomplete one.

\subsection{Content Guideline Violation Detection}\label{ContentViolation}

\subsubsection{Content safety violations}
After checking toxic content with the Perspective tool in skill outputs, we found one skill named ``Wired life hacks'' that outputs ``Anything else you would like to know, mother f*ckers?'' and ``you f*cked up! you f*cked up! you f*cked up!''. Such outputs can be harmful to users. Next we checked whether any skill asks for a positive rating. \song{One skill named ``Word Cyclopedia'' says ``If you like this skill, please give this skill five star and write your valuable feedback'' in the description.}

{Skills may get functional content from a website or provide users with an audio/image as output.} 
After extracting and processing media files separately, we got 1,656 websites, 1,439 mp3 files and 932 images from skill code. For websites, we checked whether they are malicious websites and whether the content of a website has toxic content.
We didn't find any such case in our dataset. 
For the extracted audio and image files, we didn't find any violation either. 
\liao{Detecting content safety from the dynamic and changing external websites (where skills obtain data) can be challenging. However, upon revisiting and comparing the content of all websites used in skills in a six-month period, we found that the majority of these websites (85\%) didn't change their content.} 

\subsubsection{Non-compliance of invocation names}

We found 281 skills with different types of invocation name violations. Among these skills, 53\% of skills have an invocation name with only one word, such as ``greeter'', ``eva'' or ``jokes''; 4\% of them have two-word invocation names with an article or preposition, like ``the car'' or ``the helper''; 35\% of skills used an invocation name that contains Alexa-related words, such as ``play radio'', and the other 8\% skills don't use invocation name in lower case like ``AlTranslate'' or ``Ryan's note''. \song{Example invocation names with violations are provided in Table~\ref{table:invocation_name}. \ProjectName{} successfully detected invocation name violations with an accuracy of 96\% \liao{and recall of \ls{99}\%. 
\ProjectName{} incorrectly reported skills with invocation names that are one-word brand names as violations.
The false negatives are mainly due to the failure to read the skill file with errors.}}

\song{Amazon Alexa also requires that ``the invocation name should be distinctive to ensure users can enable your skill''.
Such violations of the invocation name can negatively influence user experience or be used for squatting attack~\cite{Security:2018:Squatting}. For example, one skill's invocation name is ``weather app''. When users try to invoke the ``weather'' skill with ``Alexa, open the weather app'', this skill will be invoked instead of the ``weather'' skill. For users who want to invoke a skill with commonly used invocation name, he/she first needs to find and enable that skill, so that it can be correctly invoked from multiple candidate skills. However, we found such a violation is very common in both our dataset and Amazon Alexa skills store. So, we decided not to include it in our results. We found that 202 invocation names used by 990 skills (40\%) in our dataset, and 16,020 published skills have this violation in Alexa's skills store.}

\subsubsection{Category-specific violations}
For policy violations related to different categories, we found one skill asks for user data in the Kids category. 13 skills in the Health category lack a disclaimer in their description. For example, one skill named ``UCSD Health'' in the ``Health'' category provides a description ``Ask UCSD Health to save your spot at the nearest clinic'' and it doesn't contain a disclaimer \song{which is required by Alexa}.

\hc{for each of the subsections in Section 6, it may help to have a concise summary of key (security) findings/highlights and takeaways at the end}

\subsection{Code Inconsistency Detection}\label{CodeInconsistency}

\subsubsection{A skill with a data collection request but without a slot for processing user reply}\label{lackanintent}
When a skill asks for user data, it should define an intent and slot for capturing and processing such data, such as ``NameIntent'' with slot ``name''. However, we found that 124 skills ask for user data but don't have corresponding intent and slot. 
For example, the skill ``awsFact'' tells user ``You can introduce yourself by telling me your name'', but it only contains four Amazon built-in intents.
For such a skill without proper intent and slot, if users provide data following the instruction, the skill will invoke a wrong intent or trigger built-in ``FallbackIntent'' with the reply ``Sorry, I don't know that. Please try again''. {Such an incorrect reply will no doubt influence user experience. We found that over 50 skills use the same sentence (``You can introduce yourself by telling me your name'') in the “HelpIntent” but
don't have an intent or slot for processing the data.} 

\subsubsection{A skill with a data collection slot but without data collection requests to trigger the slot}

We found 147 skills that
define a slot for data collection, but they don't have any data collection requests. 
For example, a skill named ``theStudyBar'' has an intent named ``myNameIs''. The slot defined in this intent is ``firstName'' and the sample utterances are ``\{firstName\}'' and ``my name is \{firstName\}''. When we checked the skill back-end code, there didn't exist a request asking for the user name. 
This makes the data collection activity hidden when the skill is submitted during the certification phase. However, developers can update the code 
after the skill is published, then the skill can collect user data again. 

\subsubsection{Data collection slot and intent without sample utterances}

We found 24 skills with 95 sensitive data collection slots and 40 skills with 87 intents (not for data collection) don't have any corresponding sample utterances. 
{This can be a bug or error because the interaction model doesn't know how to match a user's reply with the intent and slot. When a user provides a reply, no intent or a wrong intent will be invoked and the slot value will never be captured.} Such cases are similar to missing an intent because although the intent is defined, it will not be triggered.

For code inconsistency detection, we also manually checked all our results. For skills with data collection requests but without a
slot, the accuracy is 91\%. For skills with data collection slots but without data collection requests, the accuracy is 84\% because some data collection outputs cannot be correctly recognized due to the complexity of NLP. For detecting data collection slots and intents without sample utterances, \ProjectName{} achieves 100\% accuracy.
\liao{Since the inconsistency detection is reliant on the results from sensitive data collection activity detection, the reasons for false negatives are the same with Section~\ref{DataCollectionAnalysisResult} and
the recall rates for the three parts are \ls{82\%, 92\%, 93\%,} respectively.}



\subsection{\song{Impact of Code  Duplication}}
\label{publishedskills}

\subsubsection{\song{Impact of code snippet Copy\&Paste}}

Although we have removed the duplicate skills before the analysis, we still found many different skills with the same policy violations. For example, 67 skills output ``You can introduce yourself by telling me your name'' in their HelpIntent, which is designed to provide helpful information for users. However, they don't provide a privacy policy while collecting user data. Since the default sentence in HelpIntent is ``You can say hello to me! How can I help?'', it is possible that these skills didn't use Alexa's default template but copied the code snippet from other places for their functions.

To find such cases in skill violations, we used the copydetect~\cite{copydetect} tool
to detect the similarity between the source code files where skill violations appeared. Then, we grouped similar skills with the same violations and checked whether there exists any template skill, \song{especially skill tutorials or templates, which were potentially copied by other skills in the same group}. As a result, we found there exist 453 violations (38\%) because of copying code from other skills. Such cases were more serious in two types of violations: ``sensitive data collection in output but missing a privacy policy'' (53\% of the skills with such issue are due to the \song{code snippet copy\&paste}) and ``sensitive data collection in output but without a slot for processing it'' (67\% of the skills that violate this policy are due to the \song{code snippet copy\&paste}). As for the source of copied skills, we found the most commonly copied skills were from the GitHub accounts ``dabblelab'' and ``alexa-samples'', which lead to violations in another 87 and 65 skills, respectively. Note that ``alexa-samples'' seems to be an Alexa official account and it contains 109 skill tutorials. More details about Alexa's official accounts will be discussed in the next section.

\subsubsection{\song{Code snippet Copy\&Paste in potential Alexa's official skills.}}
After tracing the sources of copied skills, we found that the GitHub accounts ``alexa-samples'' , ``alexa'', ``alexa-dev-hub'', ``alexa-labs'' and ``aws-samples'' (which are potential Amazon Alexa's accounts on GitHub), published skills that were often copied by other skills and they lead to violations in tens of skills. Although these Alexa-provided skills aim to provide tutorials for developers to learn how to develop skills, a few issues in these skills could lead to violations in other skills and provide bad examples for third-party developers. To understand whether these official skills' code has violations, we checked all their skills and found these accounts don't provide a good example for others. The account ``alexa-samples'' published 125 skills, in which 20 skills are supposed to provide a privacy policy but only 1 skill provides the privacy policy. This account influenced 62 other skills. What's more, instead of leaving a blank for the privacy policy link, as shown in Listing 2, some official skills don't even have the ``privacyAndCompliance'' attribute in the manifest file, which makes developers not notice there should be a privacy policy link. Another issue is that the tutorial skills are in different standards regarding policy compliance. For a template named ``zero to hero'', which includes 10 different versions of skill templates, only the tenth skill provides a privacy policy while the others do not, although all ten skills ask for the same user data. 
In total, we found 42 official skills with violations and they influenced 71 other skills with 81 policy violations. 

\subsection{Performance and Responsible Disclosure}\label{Latency}

\subsubsection{Performance}
\liao{Table~\ref{table:performance} summarizes the performance of each component in \ProjectName{} and Table~\ref{table:results} summarizes our detection results. 
\ls{For each component, we manually checked all the detected results to get the accuracy of \ProjectName{}. In addition, we}
also randomly selected \songl{100} skills from our dataset and manually labeled them to evaluate the 
\songl{overall performance of}
\ProjectName{}. The results show that \ProjectName{} achieves \ls{overall 90}\% precision and \ls{87}\% recall.}

\begin{table}[!h]
	\centering
	\vspace{5pt}
	\resizebox{8.5cm}{!}{
			\begin{tabular}{  
			>{\centering\arraybackslash}m{1.7cm}
			>{\arraybackslash}m{5.7cm} 
            >{\centering\arraybackslash}m{1.0cm} 
            >{\centering\arraybackslash}m{1.0cm} 
			}
				\Xhline{5\arrayrulewidth}
				\bf Analysis&
				\bf{\ProjectName{} Component}&  {\bf Accuracy}	& \liao{Recall}
				\\
				\hline
				\multirow{3}{*}{\shortstack{Data \\Collection\\ Activity}} & Sensitive data collection request detection & 98\% & \ls{84}\%\\
				\cline{2-4}
				& Sensitive data collection slot detection & 99.7\% & \ls{93}\% \\
				\cline{2-4}
				& Sensitive data collection permission detection & 100\% & \ls{91}\% \\
				\hline
				\multirow{2}{*}{\shortstack{Data Flow \\Analysis}}&Slot data flow analysis & 95\% & - \\
				 \cline{2-4}
				& Permission data flow analysis & 89\% & - \\
				\hline	\multirow{2}{*}{\shortstack{\vspace{-2pt} Privacy\\ Violation}} & Missing a privacy policy & 100\% & \ls{86}\% \\
				\cline{2-4}
				& Having an incomplete privacy policy & 93\% & \ls{86}\% \\
				\hline
				Content &Invocation name violation & 96\% & \ls{99}\% \\
				\hline
				\multirow{4}{*}{\shortstack{Code\\ Inconsistency}} & Data collection request but missing a slot & 91\% & \ls{82}\%\\
				\cline{2-4}
				& Data collection slot but missing a request & 84\% & \ls{92}\% \\
				\cline{2-4}
				& Data collection slot but missing an utterance & 100\% & 93\% \\
				\cline{2-4}
				& Intent but missing an utterance & 100\%  & 93\%\\
                \hline
                \rowcolor[gray]{.9} \multicolumn{2}{c}{Overall performance of \ProjectName{}} & 90\% & \ls{87}\%\\

				\Xhline{5\arrayrulewidth}
	\end{tabular}
	}
	\caption{Performance of different components in \ProjectName{}.}
	\label{table:performance}
	\vspace{-25pt}
\end{table}

\subsubsection{Latency}
\begin{figure}[!h]
	\begin{center} 
	\vspace{-5pt}
		\includegraphics[width=0.35\textwidth]{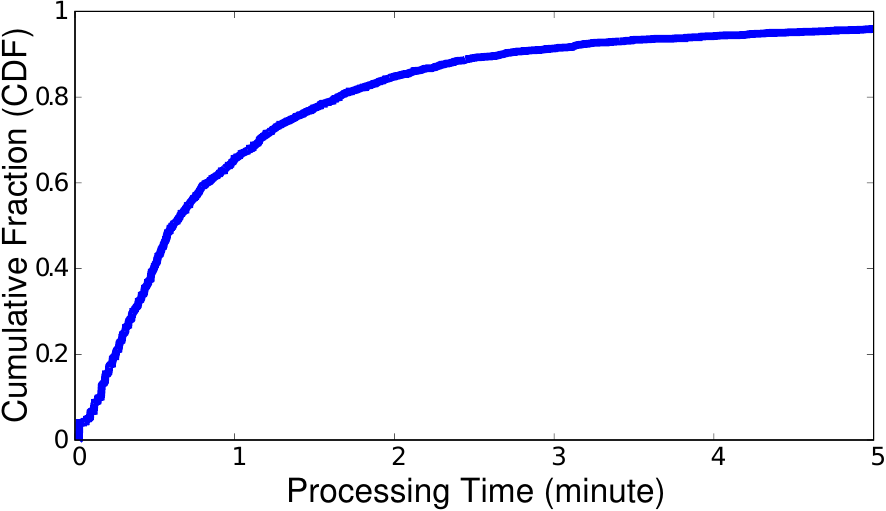}
	\end{center} \vspace{-12pt}
	\caption{Latency performance of \ProjectName{}.}
 \vspace{-10pt}
	\label{latency}
\end{figure}

\liao{For each skill, \ProjectName{} needs only 75s on average for scanning all source files, performing the policy violation detection, and reporting analysis results. Figure~\ref{latency} shows the Cumulative Distribution Fraction (CDF) of \ProjectName{}'s latency performance. Over 85\% of skills need less than two minutes for a comprehensive detection. 
Compared to SkillDetective~\cite{SkillDetective:2022}, which uses a dynamic testing method that needs to wait for skill responses and takes around 10 minutes to test one skill, \ProjectName{} has a significantly lower time cost.
\songl{The majority of time consumed by \ProjectName{} is on toxic content detection (sending outputs to Perspective) and downloading audio, images, and websites.}
}

\subsubsection{Responsible disclosure}
\label{responsible}
\liao{We performed a responsible disclosure process with several steps. 1) We reported the identified issues in Alexa's sample code to the Amazon Alexa team. 2) We notified 83 developers who provided email addresses on their GitHub pages about the potential policy violations in their code. 3) We created new issues on the GitHub repositories to notify the developers who had multiple problematic skills but didn't provide an email address.
It is worth mentioning that we received feedback from 6 developers \ls{(among the 83 developers we reached out to)}. They expressed appreciation for our work and acknowledged that such a static analysis tool is helpful in facilitating their development of policy-compliant skills. \ls{They} also mentioned the reasons for the policy violations, such as being unaware of policies, using a template that does not have a privacy policy or developing skills for learning purposes.}
\ls{Three respondents mentioned that they were impressed by our \ProjectName{} tool. Two respondents planned to fix the issues and one had already fixed the issue. Two respondents explained that the skills were just developed for learning purposes.}

\section{Discussion}

\subsection{Potential Impact of Problematic Skills }
\ls{
Problematic skills could have a negative impact on users, VPA platforms, and developers.
1) \textit{End-users are concerned about policy violations in skills}. 
\songl{We collected a user review dataset from the skills store and 
we did find negative reviews talking about the policy violations in skills that impacted their experience.}
2,383 users complained about the data collection and 107 users mentioned that skills repeatedly asked for a positive rating in reviews. 
2) \textit{VPA platforms are concerned about policy violations in skills}. 
The Amazon Alexa platform enforces a certification process to ensure that each skill meets the required content and privacy requirements before it becomes publicly available. 
\songl{
According to recent works~\cite{Jide:2022:WWW, edu2021skillvet,SkillDetective:2022}, 
after researchers reported the policy-violating skills
in the skills store 
to the Amazon Alexa team, VPA platforms would remove these skills from the store.}
Out of 93 skills in our dataset published in the Alexa skills store, \ProjectName{} detected 14 problematic skills that either missed a privacy policy or had an invalid invocation name. Interestingly, the policy violations in these skills were fixed in their published versions. It is possible they failed to pass Amazon's certification process with policy violations and then the developers had to resolve the issues for publishing them. 
3) \textit{Developers are concerned about policy violations in skills}. Since policy-violating skills may be rejected at the certification phase and developers need to fix issues in skills and wait several days to get feedback from the certification team, policy violations in skills inevitably slow down the skill development and publishing. }

\subsection{Limitations}


Despite the limited number of skill packages from GitHub, \ProjectName{} is able to identify 1,328 violations in 786 policy-noncompliance skills. 4 policy-violating skills have been published on the skills store without fixing them \liao{(2 skills have an incomplete privacy policy and 2 skills in the Health category lack a disclaimer)}. However, \ProjectName{} has several limitations. First, for the data flow analysis, more types of taint sources and sinks can be considered. For example, a skill may collect users' email addresses and send emails to users in the back-end code. Account linking~\cite{SkillAccountLinking} can be another way of user data leakage and we will consider that in our future research. 
Second, the current design of \ProjectName{} does not cover all Alexa's privacy requirements and content guidelines. In addition, the policy violation detection performance can be further improved by using advanced machine learning techniques. However, it is non-trivial to collect high-quality datasets for model training. 
Third, \ProjectName{} only focuses on the Amazon Alexa platform, which is the most popular VPA platform. 
We plan to extend \ProjectName{} to analyze Google Actions as our future work. 
\liao{Forth, most of the skills in our dataset were not published in the skills store and they need to pass a certification process before being published. As a result, the skills in our dataset may have more issues than the skills in the store. However, we don’t claim the identified issues in our results are representative of actual skills in production. Rather, the main purpose of our evaluation in Section~\ref{evaluation} is to demonstrate the efficiency of \ProjectName{} in finding policy violations given skill code.}
Since \ProjectName{} is designed for facilitating third-party developers in developing policy-compliant skills, it is important to evaluate the developer acceptance rate of \ProjectName{}. We also plan to conduct user studies to get feedback from skill developers about the usability of \ProjectName{}.

\section{Related Work}

There has been an increasing number of studies on VPA security and privacy~\cite{Edu:2021:Surveys, Chen:2022:Survey, Peng:2022:PIEEE, yan2020surfingattack}. 
The majority of research efforts have been undertaken to identify various acoustic-based attacks (\eg, out-of-band signal attacks and adversarial example attacks) against the Automatic Speech Recognition (ASR) modules in VPA systems~\cite{SoK:Hadi:2021, chen2019real, Carlini:Security:2016, yan2020surfingattack} and the corresponding defenses in mitigating these attacks~\cite{Zhang:2017:DIV, ahmed2022towards, meng2022your, chen2021fakewake}.
As hundreds of thousands of skills have been published in VPA platforms, the security of skills has attracted attention from the research community.
Since attacks that exploit skills' vulnerabilities can be launched remotely, they could potentially be more powerful than acoustic-based attacks~\cite{zhang:NDSS:2019, wang2022ghosttalk, esposito2022alexa}.
In this section, we briefly discuss recent research on identifying problematic (\eg, privacy-invasive) skills in Amazon Alexa and Google Assistant platforms.

Kumar~\etal~\cite{Security:2018:Squatting} presented the skill squatting attack and they found 381 pairs of skills were likely to be squatted.
Lentzsch~\etal~\cite{Lentzsch:NDSS:2021} discovered 262 skills with permissions provide an incomplete privacy policy. 
In~\cite{Jide:2022:WWW}, the authors reported 675 skills that request permission data have privacy issues about privacy policies. 
\song{SkillVet~\cite{edu2021skillvet} presented a machine learning based method for checking data practices and privacy issues.}
SkillExplorer~\cite{SkillExplorer:2020} 
tested 28,904 Amazon skills and identified 1,141 skills requesting users to provide personal information without disclosing in their privacy policies.
VerHealth~\cite{Shezan:VerHealth:2020} analyzed 813 health-related skills on the Amazon Alexa platform. \liao{Vitas~\cite{li2022vitas} interacted with 41,581 skills and found that 51\% of skills suffered from problems such as unexpected exit/start and privacy violation.}
SkillDetective~\cite{SkillDetective:2022} identified 6,079 policy-violating skills in the current skills stores by evaluating skills' conformity to more than 50 different policy requirements.
Many research efforts have been undertaken to study user concerns (human factors) about
the security and privacy of VPA devices~\cite{ Chung:2017:Computer, Serena:2018:HCI, Ammari:2019:MSI, SOUPS:2019}.
Malkin~\etal~\cite{malkin2022runtime} conducted a user study about how people react to the runtime permission. 
Sharma~\etal~\cite{sharma2022understanding} focused on Google Assistant and showed that most participants have superficial knowledge about the data collected by the platform. 
Liu~\etal~\cite{liu2022effects} studied 214 participants about whether they consider privacy permissions while installing apps. Sabir~\etal~\cite{sabir2022hey} did a user study to analyze users' awareness of third-party skills.  However, none of them conducted user study to understand skill developers' perceptions and practices regarding policy compliance in skill development.

\liao{{\bf Distinction from existing works.} \ProjectName{} distincts itself from existing policy violation detection works (including SkillExplorer~\cite{SkillExplorer:2020}, VerHealth~\cite{Shezan:VerHealth:2020}, Vitas~\cite{li2022vitas} and SkillDetective~\cite{SkillDetective:2022}) in three ways.
First, different from recent skill testing
tools that utilize dynamic analysis and can only detect violations exercised by run-time inputs, \ProjectName{} is the first-of-its-kind static analysis tool and it detects more possible violations/inconsistencies in the code. 
Second, SkillScanner identifies policy violations of skills developed by inexperienced benign developers in the development phase while existing works detect policy violations in the post-deployment phase.
Third, \ProjectName{} detects more types of violations existing in the skill code (such as data storage, over-privileged data, and code inconsistency) that existing works can't find. For example, compared to SkillDetective, \ProjectName{} detects 12 new types of violations. Even for the 8 types of violations that SkillDetective can find, such as skill output and content safety, \ProjectName{} can find more potential outputs that might not appear in dynamic testing. As a result, 934/1328 (70\%) of violations are new and can't be found by SkillDetective.
}

\section{Conclusion}

In this work, we first conducted a user study to understand the gap between VPA's policy requirements and skill developers' practices. 
Informed by our user study results, we
designed and implemented the first static analysis tool named \ProjectName{}, which helps 
developers automatically identify potential policy violations in skills at the development phase. To evaluate the performance of \ProjectName{}, we collected 2,451 open source skills from GitHub, and conducted a comprehensive analysis of these skills using \ProjectName{}. \ProjectName{} effectively identified 1,328 violations among 786 skills. 694 of them are about privacy non-compliance and 298 of them violate content guidelines imposed by the Amazon Alexa platform. \song{We found that 32\% of policy violations are introduced through code copy and paste. The policy violations in Alexa's official accounts led to 81 policy violations in other skills.} As our future work, we plan to conduct user studies to evaluate the usability (\eg, acceptance and user-friendliness) of \ProjectName{} by skill developers.  

\section*{Acknowledgment}

The work of L. Cheng is supported by National Science Foundation (NSF) under the Grant No. 2239605, 2228616 and 2114920.
The work of H. Hu is supported by NSF under the Grant No. 2228617, 2120369, 2129164, and 2114982.
The work of L. Guo is supported by NSF under grant IIS-1949640, CNS-2008049, and CCF-2312616.

\begin{table*}[!htbp]
	\centering
	\resizebox{18cm}{!}{
		{ 
			\begin{tabular}{
			|>{\centering\arraybackslash}m{1.8cm}| >{\centering\arraybackslash}m{4cm} |
			>{\centering\arraybackslash}m{9cm} |
			>{\centering\arraybackslash}m{3cm}| >{\centering\arraybackslash}m{2cm}| 
					}
			\Xhline{3\arrayrulewidth}
			
			\rowcolor[gray]{.9}& Problem & Detailed policy & Data source & \# of skills with policy violation
			\\
			\cline{1-5}
			
			\multirow{13}{*}{\shortstack{Privacy \\ Violations}} & 
			
			\multirow{4}{*}{\shortstack{Data collection/storage but\\missing a privacy policy}} & \multirow{4}{*}{ \shortstack{Provide a legally adequate privacy notice that will be displayed\\ to end users on your skill's detail page. }} & Output & 240
			\\
			\cline{4-5}
			& & &  Permission & 94
			\\
			\cline{4-5}
			& & &  Database & 126
			\\
			\cline{4-5}
			& & &  Disclosure to Alexa & 4
			\\
			\cline{4-5}
			
			\cline{2-5}

			& \multirow{4}{*}{\shortstack{Data collection/storage but\\ having an incomplete privacy \\policy}} &  \multirow{4}{*}{\shortstack{Ensure that your collection and use of that information\\ complies with your privacy notice and all applicable laws.}}  & Output & 19
			\\
			\cline{4-5}
			& & & Permission & 23
			\\
			\cline{4-5}
			& & & Database & 10
			\\
			\cline{4-5}
			& & & Disclosure to Alexa & 19
			\\
			\cline{2-5}				
			
			& Over-privileged data requests	& Collect and use the data only if it is required to support and improve the features and services your skill provides. & Permission & 42
			\\
			\cline{2-5}
			
			& Not-asked permission usage	& Bug & Permission & 18
			\\
			\cline{2-5}			
		    
			& \shortstack{Incorrect data collection\\ disclosure to Alexa}	& Wrong answer to ``Does this Alexa skill collect users' personal information?'' & Disclosure to Alexa & 99
			\\
			\cline{1-5}	
			
			\multirow{9}{*}{\shortstack{Violations of \\ Content \\Guidelines}} & Content safety & Contains excessive profanity.  & Output & 1
			\\
			\cline{2-5}
			& Asking for positive rating & Can not explicitly requests that users leave a positive rating of the skill. & Description & 1
   \\
			\cline{2-5}
			& Invocation name requirements & Does not adhere to Amazon Invocation Name Requirements. Please consult these requirements to ensure your skill is compliant with our invocation name policies. & Invocation name & 281
			\\
   
			\cline{2-5}
			& Kid category policy & It doesn't collect any personal information from end users. & Output & 1
			\\
			\cline{2-5}
			& Health category policy & A skill that provides health-related information, news, facts or tips must include a disclaimer in the skill description stating that the skill is not a substitute for professional medical advice. & Description & 13
			\\
			\cline{1-5}
			
			\multirow{6}{*}{\shortstack{Code\\ Inconsistency}} & Data collection \liao{request} but missing a \liao{slot} & Bug  & Output & 124
			\\
			\cline{2-5}
			& Data collection slot but missing a request & Vulnerability & Intent & 147
			\\
			\cline{2-5}
			& Data collection slot but missing an \liao{utterance} & Bug & Slot & 24
			\\
			\cline{2-5}
			& Intent but missing an \liao{utterance} & Bug & Intent & 40
			\\

            \Xhline{3\arrayrulewidth}
			\end{tabular}
		}
	}
	\caption{\song{Policy violations and code inconsistency in skill code and detailed policies they violate. We have removed the false positives by our manual verification.}}
        \vspace{-20pt}
	\label{table:results}
\end{table*}

\balance
\bibliographystyle{plain}
\bibliography{ref.bib}


\end{document}